\documentclass[prl,twocolumn,lengthcheck,showpacs]{revtex4-1}
\newif\ifwithsupp
\withsupptrue   
\def\coloronline{\ifwithsupp\else(Color online)\ \fi}

\usepackage{amsmath}
\usepackage{amssymb}
\usepackage{graphicx}
\usepackage{hyperref}
\usepackage{color}

\def\idty{{\leavevmode\rm 1\mkern -5.4mu I}} 

\def\Cx{{\mathbb C}}
\def\Ir{{\mathbb Z}}

\def\norm #1{\Vert #1\Vert}

\def\ket #1{\vert #1\rangle}
\def\braket #1#2{\langle #1 \vert #2\rangle}

\def\abs#1{\vert#1\vert}

\def\Ccoll{{\Gamma}} 
\def\HH{{\mathcal H}}
\def\FF{{\mathbb F}}

\def\citepaps#1{\ifwithsupp Appendix~#1\else\cite[Part #1]{epaps}\fi}
\def\papsection#1#2{\ifwithsupp \section*{Appendix~#1: #2}\else\section*{Part~#1: #2}\fi}

\begin{document}

\title{Bound Molecules in an Interacting Quantum Walk}

\author{Andre Ahlbrecht$^{1}$}
\author{Andrea Alberti$^{2}$}
\author{Dieter Meschede$^{2}$}
\author{Volkher B. Scholz$^{1}$}
\author{Albert H. Werner$^{1}$}
\author{Reinhard F. Werner$^{1}$}
\affiliation{$^{1}$Institut f\"ur Theoretische Physik, Leibniz Universit\"at Hannover, Appelstr. 2, 30167 Hannover, Germany\\
$^{2}$Institut f\"ur Angewandte Physik, Universit\"at Bonn, Wegelerstr. 8, 53115 Bonn, Germany}

\begin{abstract}
We investigate a system of two atoms in an optical lattice, performing a quantum walk by state-dependent shift operations and a coin operation acting on the internal states. The atoms interact, e.g., by cold collisions, whenever they are in the same potential well of the lattice. Under such conditions they typically develop a bound state, so that the two atoms effectively perform a quantum walk together, rarely moving further from each other than a few lattice sites. The theoretical analysis is based on a theory of quantum walks with a point defect, applied to the difference variable. We also discuss the feasibility of an experimental realization in existing quantum walk experiments.
\end{abstract}


\maketitle
\ifwithsupp\else\nocite{epaps}\fi 

Quantum walks are the discrete-time and discrete-space analog of free quantum particle motion. They exhibit many of the well-known features of quantum dynamics: In the translation invariant case their motion is governed by a dispersion relation, from which the position distribution at large times is obtained as the distribution of group velocity. By modifying the coin operation of the quantum walk on a small number of sites, one can create the analog of potentials, with non-trivial scattering cross sections and bound states. In this paper we consider the possibility of an interaction between two walking particles, which becomes effective whenever the particles occupy the same site. In the simplest case, the Hadamard quantum walk, the interaction consists of just an additional phase. We show that with such interaction one can expect the formation of molecules, with all the features known from ordinary quantum dynamics of interacting two-particle systems: The two atoms stay together forever, performing a quantum walk in its own right.
The distance between the particles remains bounded in the sense that they occupy nearby lattice sites with high probability, with estimates valid for all times.

Quantum walks have recently been demonstrated experimentally in setups based on controlled neutral atoms \cite{Bonn}, ions  \cite{Schaetz,Blatt}, individual photons \cite{Broome}, wave guides \cite{Peruzzo:2010p3068,Silberberg08} or light pulses \cite{Silberhorn}. While all these experiments show the expected one-particle dynamics, their ability to demonstrate the type of interaction we describe differs very much. In some realizations interaction is very natural if not unavoidable: when two atoms sit in the same potential well of an optical lattice they are expected to pick up a collision phase depending on the duration of the contact. In spite of the fact that the atom-atom interaction would be much too weak for chemical binding in free space, the lattice creates a complex interference effect, which nevertheless leads to binding. The relevant conditions and parameters for an implementation of molecules of two atoms walking in a 1D optical lattice are discussed below. In a suitable system molecules are easy to prepare by starting two particles on the same site. The initial wave function then typically has a good overlap with the bound state, and the group velocity of unbound pairs is higher than that of molecules. Hence one can just wait for the unbound atom pairs to move away. Our conclusion is that an experiment should be possible in the near future, perhaps even by extending an existing experiment.

Our present work was partly inspired by analogous work in the continuous-time case \cite{Zoller}, in which the motion between neighboring sites is given by tunneling. The spectral structure of continuous-time lattice dynamics is indeed analogous to that of discrete-time walks, except for the fact that in discrete time the spectrum of the unitary one-step operator is on the unit circle, and in the continuous-time case the spectrum of the Hamiltonian is on the real line, so it makes sense to speak of high energies and low energies. In contrast, for discrete-time quantum dynamics there is no distinction between repulsive and attractive interaction potentials. Hence, in contrast to \cite{Zoller}, we cannot speak of ``repulsively bound'' molecules. In the discrete time case, a recent study of correlations in two-particle quantum walks \cite{Stefanak}, numerically found enhanced correlations related to the molecule states studied in this paper.

\paragraph{Free Quantum Walks.---} \hspace{-1em}
Let us begin by describing quantum walks of a single particle. A walk runs on an $s$-dimensional lattice, which we take as the cubic lattice $\Ir^s$. The walking particles generally have an additional quantum degree of freedom, described in a $d$-dimensional Hilbert space $\Cx^d$. We can think of this as the internal states, or a ``coin'' degree of freedom, which can be ``flipped'' by a suitable unitary operation. The basis states of the system Hilbert space are thus $\ket{x,\alpha}$ with position $x\in\Ir^s$ and internal state label $\alpha$. The dynamical time step is given by a unitary operator $W$, which commutes with lattice translations. A crucial condition is locality: $W$ has non-zero matrix elements only between states on lattice sites closer together than some fixed finite distance, called the neighborhood size of the walk. Since $W$ commutes with translation it is convenient to partially diagonalize this operator by Fourier transform, i.e., to look at the walk in momentum space. In the momentum representation wave functions depend on $p\in[-\pi,\pi]^s$, i.e., the Brillouin zone of the lattice, with $\psi(p)\in\Cx^d$.  The walk in this representation acts by multiplying the wave function at $p$ with a unitary $d\times d$-matrix $W(p)$. The locality condition then becomes the statement that each entry must be a polynomial in the variables $e^{\pm ip_k},\ k=1,\ldots,s$. The standard example is the so-called Hadamard walk in one dimension ($s=1$), for which
\begin{equation}\label{hadamard}
    W_H(p)=S(p)C=\left(\begin{array}{cc}e^{ip}&0\\0&e^{-ip}\end{array}\right)
         \frac1{\sqrt2}\left(\begin{array}{cc}1&1\\1&-1\end{array}\right)
\end{equation}
Here the first factor is usually called the (internal state-dependent) shift and the second the coin. In one lattice dimension a decomposition into (usually several) such factors is always possible, but in higher dimension this is an open question. There is no intrinsic connection between the space dimension $s$ and the internal state space dimension $d$. In particular, it is not necessary to choose a different coin for each dimension (which would give $d=2^s$ \cite{Konno}).

For a non-interacting walk of two distinguishable particles, we just take $W_2(p_1,p_2)=W_1(p_1)\otimes W_1(p_2)$ where $p_1$ and $p_2$ are the momentum coordinates of the two particles. Apart from a doubled lattice dimension ($s'=2s$) and a squared dimension of the internal state space ($d'=d^2$), this is a walk exactly as described above. Obviously, since $W_1\otimes W_1$ commutes with the particle permutation, we may restrict the two particle walk to either the Bose or Fermi subspace. The influence of entangled initial states on the position distribution of the non-interacting walk has been studied theoretically in \cite{OmaBose}, and experimentally with photons in \cite{Peruzzo:2010p3068}.

%
%
\paragraph{Interacting Quantum Walks .---}\hspace{-1em}
To make a two-particle walk interacting we need to introduce a dependence on the particle coordinates. The general way to introduce such modifications is to define a new transition operator as $W'=W_2C$, where $W_2$ is the unperturbed, fully translation invariant two particle walk, and $C$ is a space dependent coin operation, i.e., a unitary which acts (in a possibly different way) on the internal degree of freedom at each site.  We choose an operator $C$, which differs from the identity operator $\idty$ only when the two particles are on the same lattice site ($x_1=x_2$), in which case it acts on the joint coin space as some fixed operator $\Ccoll$. We denote by $N$ the projection operator onto the set of collision points, i.e., the diagonal in the $x_1,x_2$-diagram.
The overall interacting walk operator is
\begin{equation}\label{W2}
    W_\Ccoll=(W_1\otimes W_1)\bigl( (\idty-N)+ \Ccoll N\bigr).
\end{equation}
Note that $N$ only acts on the translation degrees of freedom and $\Ccoll$ acts on the coin space, so these operators commute and the second factor is unitary. Since $\Ccoll$ is taken to be the same for every collision point, we thus break the separate translation invariance, but retain the invariance of the walk under joint translations. Therefore, the total momentum $p_1+p_2$ is conserved by $W_\Ccoll$.

We will develop the basic theory for such walks quite generally, for any $\Ccoll$ in any coin dimension $d$ and also any lattice dimension (see \citepaps A for the salient conditions). However, for illustration we will focus especially on the simplest case, which we call a walk with {\it singlet collisions}. This is defined by having the coin space at the collision points one-dimensional, so the coin acts just by a phase $\Ccoll=\gamma\idty$. In the Fermi case, and when the single particle coin space is two dimensional as for the Hadamard walk, this condition is automatically fulfilled even for the free walk. In the Bose case of the Hadamard walk, particles can only end up on the same site if they come from opposite directions, i.e., in a ``symmetric singlet'' $(\ket{\!\uparrow\downarrow}+\ket{\!\downarrow\uparrow})/\sqrt2$. If the combination of free coin and collision coin leave this state invariant, we are again in the singlet collision case.

In order to define the interaction for more than two particles, or for a quantum lattice gas of such particles, more work is required. We describe this in the supplement \citepaps B, mostly for the case of singlet collisions.

\begin{figure}[ht]
  \includegraphics[width=8.2cm]{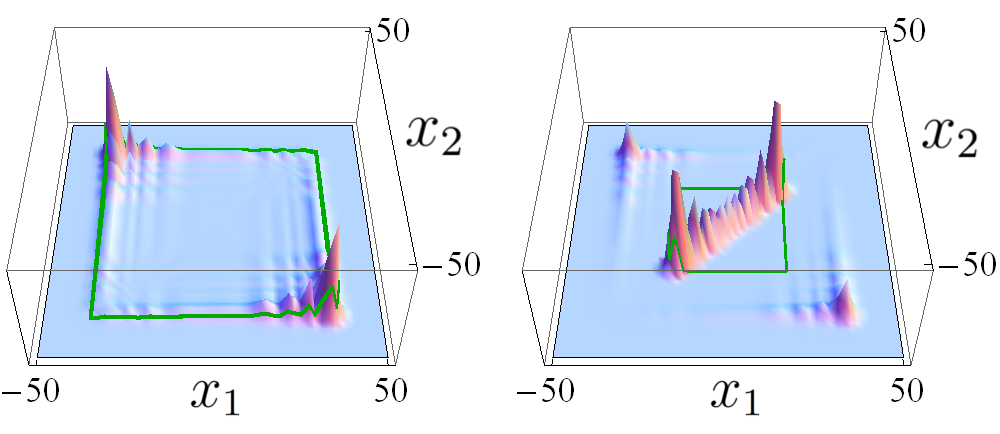}
  \caption{\coloronline Joint probability distribution of two particles after $t=50$ steps of a Hadamard walk (antisymmetric subspace). Left panel: without interaction ($\gamma=1$); Right panel: interaction with collision phase $\gamma=-1$. The embedded (green) squares are for comparison with the theoretically calculated peak velocity of the free Hadamard walk (left panel) and the peak velocity of the effective walk of molecules after Eq.~\eqref{velo}. }
  \label{fig:positions}
\end{figure}

\paragraph{Numerical Phenomenology.---}\hspace{-1em}
We start the walk \eqref{W2} from the initial state in which both particles are at the origin, and their joint internal state is antisymmetric. Since the Fermi subspace is invariant under $W_\Ccoll$, all subsequent collisions will also be of singlet type. Then by multiplying the initial vector $t$ times with \eqref{W2}, and taking the modulus square with respect to the internal states, we get the joint position distribution of the two particles at time $t$. The result is shown in Fig.~\ref{fig:positions}, once for the non-interacting case ($\gamma=1$), and once for the interacting case with value $\gamma=-1$.
Note that due to Fermi symmetry both diagrams are symmetric with respect to reflection at the diagonal. The peaks near $x_1=-x_2$ are expected from the theory of the free Hadamard walk \cite{timerandom}, and are near $\pm\sqrt2$. The striking concentration of probability near the diagonal is the hallmark of the bound state. The width of this distribution in the off-diagonal direction remains constant in time, whereas along the diagonal it shows the characteristic behavior of a 1D walk. This will be proved below, and the peak velocity of this walk will be determined via Eq.~\eqref{velo} to be $1/3$. For comparison, squares with edges at $\pm1/\sqrt2$ and $\pm1/3$ are drawn in Fig.~\ref{fig:positions}.

\begin{figure}[ht]
  \includegraphics[width=8.2cm]{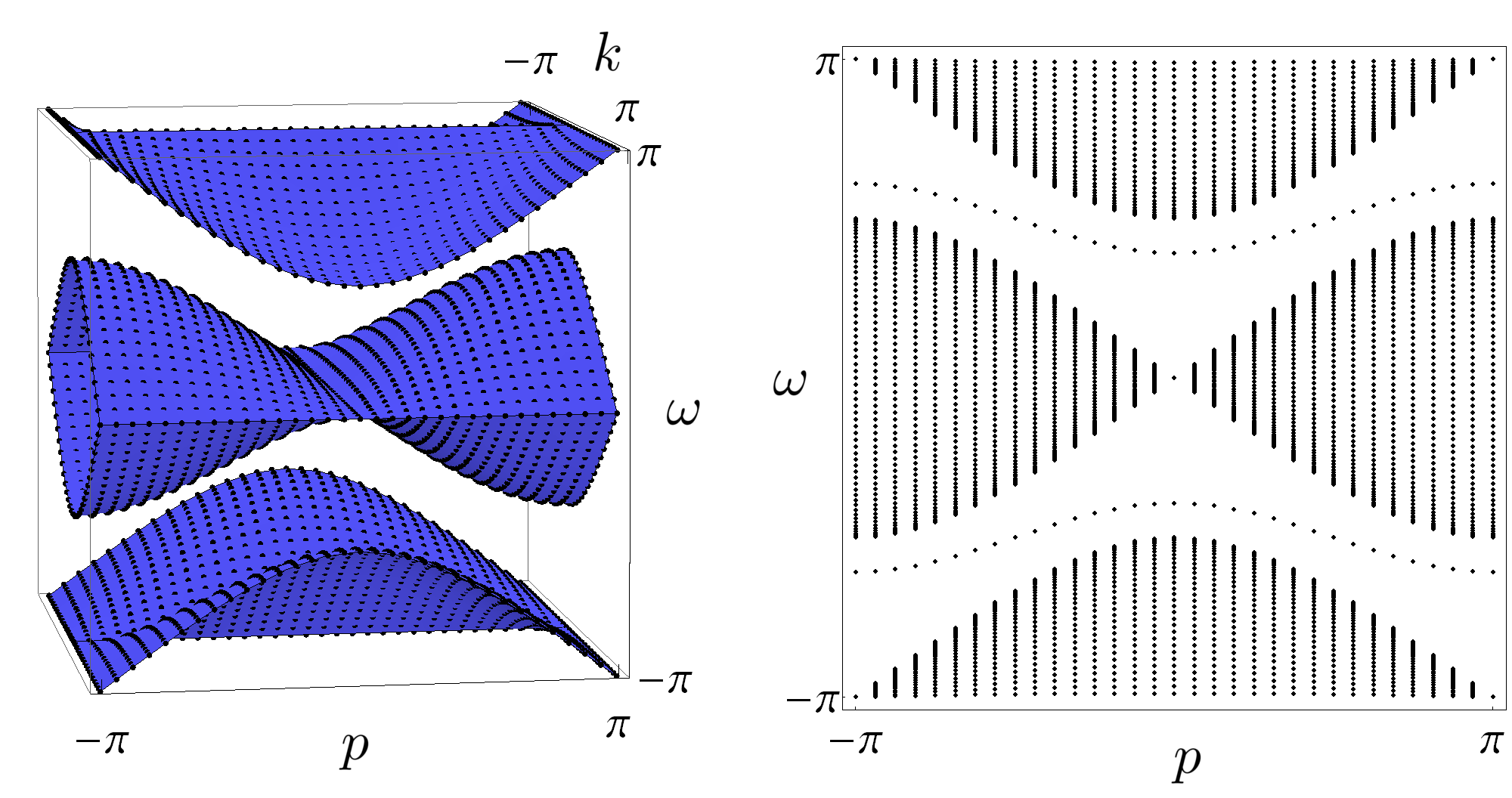}
  \caption{\coloronline Spectrum of the interacting walk operator for two particles on a ring (28 sites). Left Panel: The non-interacting case: The relative momentum is then a good quantum number, and is represented as a third direction, orthogonal to the paper plane. For large rings the points fill the indicated surfaces. Right Panel: The interacting case ($\gamma=-1$). The bands are very nearly equal to the projection of the left-panel figure onto the $(p,\omega)$-plane. The additional feature introduced by interaction is the chain of simple eigenvalues in the band gap, i.e., the molecule states.} \label{fig:spec1}
\end{figure}

Turning from space and time diagrams to momentum and energy, let us consider the spectral properties of the operator \eqref{W2}. Since it commutes with joint translations, we can diagonalize it together with total momentum $p$. For each value of $p$ we get an operator $W_\Ccoll(p)$, whose eigenvalues we can compute. To treat this numerically, we close the system to a ring, which also discretizes $p$. The eigenvalues of $W_\Ccoll(p)$, as a function of $p$, are shown in Fig.~\ref{fig:spec1}. Again, one part of this diagram is expected from the non-interacting case. Indeed, in that case $W_\Ccoll(p) = W_2(p)$ also commutes with translations in the difference variable $(x_1-x_2)$, which implies absolutely continuous spectrum for $W_\Ccoll(p)$. These bands are seen in discretized form in Fig.~\ref{fig:spec1}, and the indication of continuous bands is that the spectral density increases with the size of the ring. In addition, however, we see a single line in the gap between the bands. This is the bound state.

\paragraph{Analysis: quantum walks w ith a point defect.---}\hspace{-1em}
We first rewrite the problem in center of mass and relative coordinates. In momentum space this corresponds to the passage from the variables $(p_1,p_2)$ to
$p=p_1+p_2$ and $k=(p_1-p_2)/2$. Since $p$ is conserved we can fix $p$ and study the operator $W_\Ccoll(p)$ as an operator acting on the $k$ coordinate and the internal states. In this problem (for which $p$ is just an external parameter) the interaction appears as a modification of a free walk acting only at the origin.
The analysis of point defects in a general quantum walk is of some interest in its own right. We therefore drop the parameter $p$ for the moment, and consider this general problem. So let
\begin{equation}
W(k) = W_2(p,k) = W_1\left(\frac{p}{2} + k\right) \otimes W_1\left(\frac{p}{2} - k\right)
\end{equation}
denote the free walk matrix in relative momentum space, which corresponds, for each value of $p$, to a single particle 1D walk with a four-dimensional coin space. We thus have to analyze the family of 1D quantum walks $W_\Gamma=W_2\bigl(\idty-N+N \Ccoll\bigr)$, where $N$ is the projection onto the subspace at the origin, and $\Ccoll$ is the coin modification at the origin.

Since $N$ is a finite rank projection, the bands, i.e., the continuous spectrum of $W_\Ccoll$, are the same as those of $W_2$ \cite[Thm.IV.5.35]{FA}. To find the eigenvalues, suppose that $\Psi(k)$ is the momentum space representation of an eigenvector with eigenvalue $z$. Then $W(k)(\idty+(\Ccoll-\idty)N)\Psi(k)=z\Psi(k)$. Note that $N\Psi$ is just the value of $\Psi$ at the origin in position representation, so $N\Psi(k)=\psi$ is a function independent of $k$. This $\psi$ determines the whole function $\Psi$ by
\begin{equation}\label{resolve}
    \Psi(k)=(W(k)-z)^{-1}W(k)(\idty-\Ccoll)\psi.
\end{equation}
The condition for an eigenvalue $z$ is thus that $\Psi$ is normalizable and $\psi=N\Psi$. Since $N$ is the projection onto the zero Fourier component, i.e., integration with respect to $k$, we introduce the operator
\begin{equation}\label{rr}
    R(z)=\frac1{(2\pi)^s}\int \!\!d^sk\ (W(k)-z)^{-1}\,W(k)
\end{equation}
and get the eigenvalue condition for $\psi$ in \eqref{resolve} to determine the eigenvector for eigenvalue $z$ in the form $R(z)(\idty-\Ccoll)\psi=\psi$. This can be rewritten as
\begin{equation}\label{RC}
    \Ccoll\psi=\bigl(\idty-R(z)^{-1}\bigr)\psi.
\end{equation}
Now it turns out that for $z$ not in the spectrum of $W$, the operator on the right is unitary (see \citepaps C). Therefore, the eigenvalue condition \eqref{RC} can be satisfied for any $z$ in the gap in Fig.~\ref{fig:spec1}, for a suitable collision operator $\Ccoll$.

\begin{figure}[ht]
\includegraphics[width=4.8cm]{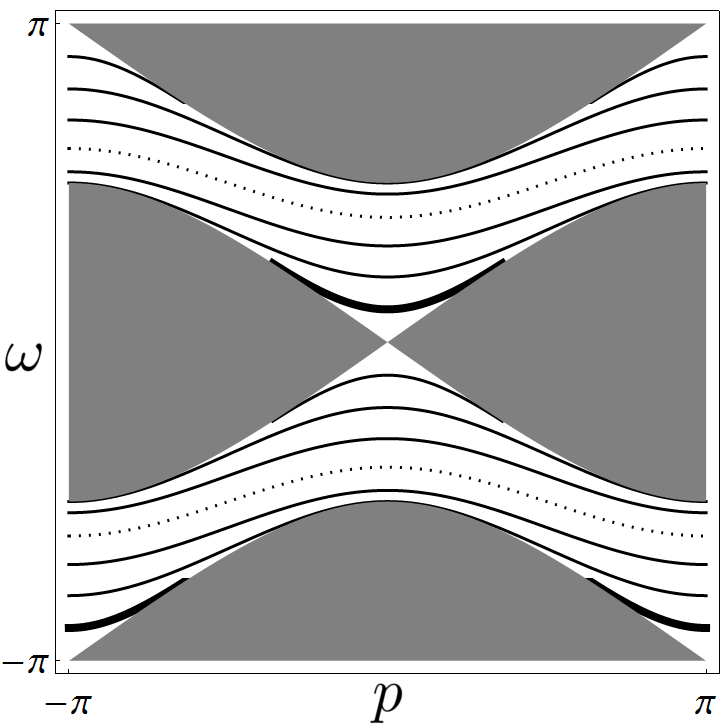}
  \caption{Dispersion relation $\omega(p)$ for the bound state. Parameter: interaction phase $g$ in steps of $\pi/4$. The thick line highlights the case $g=\pi/4$, and dots the case $g=\pi$, for comparison with Fig.\ref{fig:spec1}.}
  \label{fig:bands2D}
\end{figure}

Applying this to our standard example, the Hadamard walk with singlet collisions, the space $N$ is one-dimensional, and the integral \eqref{rr} is readily evaluated by the residue theorem, turning \eqref{RC} into a formula for the collision parameter $g$ in $\Ccoll=\gamma\idty=e^{ig}\idty$, given the total momentum $p$ and the phase  $z=e^{i\omega}$. It turns out that this relation can be solved for $\omega$, giving
\begin{eqnarray}\label{omegap}
    e^{i\omega}&=&\frac{e^{ig}}{2e^{ig}-1}\left(\cos p\pm i\sqrt{\sin^2p+4(1-\cos g)}\right)
                   \nonumber\\
               && \mskip-30mu\mbox{ provided}\ \sin\omega\cdot\sin(g-\omega)>0\,.
\end{eqnarray}
Here the constraint results from picking the correct pole for the residue evaluation in \eqref{rr}. It implies that we do not get two bound states for every momentum. Surprisingly, the branches forbidden by this constraint do not run inside the bands or off the unit circle but also in the gap. The results are shown in  Fig.~\ref{fig:bands2D}. The overlap of the initial wave function with the bound states, i.e. the probability to observe a molecule, can be computed from \eqref{rr} and \eqref{omegap}, see \citepaps D. It achieves its maximum of $\frac{2}{3}$ at $g=\pi$.

\paragraph{Coin states of the molecule.---}\hspace{-1em}
If we restrict the Hilbert space of the relative motion of the two particles just to the bound states, we have a picture resembling very much the spectrum of a single walking particle with two internal states. Is it actually the same? That is, is there a quantum walk with two internal states and {\it only nearest neighbor shifts} which reproduces exactly the walking dynamics of the molecule? For this it is only necessary to compare the spectra $\omega_\alpha(p)$, $\alpha=1,2$, because if these coincide, we can find a unitary commuting with translations and mapping one walk to the other. Since we have an explicit form of the spectrum, we can indeed identify an appropriate coin. The walk of the molecules with interaction phase $\gamma=e^{ig}$ is unitarily isomorphic to a walk with the second factor in \eqref{hadamard} replaced by
\begin{equation}\label{mwalkcoin}
    C=\frac{1}{2\gamma-1}\left(
        \begin{array}{cc}
         \gamma  & \sqrt{2} (\gamma -1) \\
          \sqrt{2} (\gamma -1) \gamma  & \gamma
         \end{array}\right).
\end{equation}
The exact form of the unitary isomorphism is determined by the bound states which are computed in \citepaps E. The walk uses both branches in \eqref{omegap}, though only ``virtually'' if one or both of them are forbidden by the constraint. It is not clear whether the bound states of general interacting quantum walks always allow such an interpretation.

\paragraph{Velocities.---}\hspace{-1em}
It is apparent from the right panel of Fig.~\ref{fig:positions} that the molecules are slower than the free particles. The square in that figure corresponds to the propagation of one site per step. The free Hadamard walk has maximal group velocity $1/\sqrt2$, corresponding to the off-diagonal peaks. The group velocity of the molecules (with $\gamma=-1$) is
\begin{equation}\label{velo}
    \frac{d\omega}{dp}=\frac{\pm\sin p}{\sqrt{4-4\cos g +\sin^2p}},
\end{equation}
with the same constraint as in \eqref{omegap}. The maximal speed (always at $p=\pi/2$) is $1/3$ for the walk with $g=\pi$, as shown  in Fig.~\ref{fig:positions}. For smaller $g$ the maximal speed according to \eqref{velo} approaches one, but becomes forbidden by the constraint. Nevertheless it is possible to design molecules, which are faster than the free atoms, see \citepaps F.

\paragraph{Prospects for experimental realization.---}\hspace{-1em}
We plan to realize walking molecules by extending an existing experiment \cite{Bonn}, in which the walk of single Cs atoms has been demonstrated. These happen to be Bosons, so the singlet collision case discussed above is not automatically realized. The internal coin states are taken as the two hyperfine states $\ket{\hspace{-3pt}\downarrow}=\ket{F=3,m_{F}=3}$, $\ket{\hspace{-3pt}\uparrow}=\ket{F=4,m_{F}=4}$, split by the atomic clock transition at 9.2 GHz. Spin-dependent shift operations work, because the up and down state see complementary circular polarization components of the optical lattice formed by counterpropagating linearly polarized waves. Internal coin operations are Rabi rotations driven by microwaves.

On-site interactions via s-wave cold collisions can be accurately described using pseudopotential methods \cite{Stock:2006p3302}. The use of Cs atoms is especially favorable, because its large triplet scattering length $a_{T}=2400\ a_{0}$ allows fast interactions \cite{Leo:2000p1959}.
Roughly, given a 3D optical lattice with isotropic harmonic confinement $\omega=2\pi\times 30$ kHz, the state $\ket{\psi_{0}}=(\ket{\!\uparrow\downarrow}+\ket{\!\uparrow\downarrow})/\sqrt2$ acquires a collisional phase shift $g=\pi$ in about 10 $\mu$s. This is just below the current step times of our experiment, so could well be incorporated. A further effect of the interaction is a level shift for $\ket{\psi_{0}}$, which may render the coin operations at 9.2 GHz ineffective at collision points, thus suppressing the transitions to $\ket{\!\uparrow\uparrow}$, $\ket{\!\downarrow\downarrow}$. In this way a singlet collision case could also be realized for Bosons.
For the collision phase to be well controlled and coherent, it is necessary to cool the motional states to the ground state not only along the axis but also in the lateral direction, in which at the moment the confinement is much weaker. We plan to enhance the lateral confinement by an additional blue-detuned doughnut-mode laser. Then  cooling to average vibrational quantum numbers $\langle n\rangle< 6 \cdot 10^{-3}$ in all three directions seems feasible, which should allow sufficiently many coherent steps to clearly demonstrate the molecule binding effect.

\paragraph{Acknowledgments.---}\hspace{-1em}
Both groups acknowledge the support by the DFG (grant Forschergruppe For 635),  and the EU projects CORNER, COQUIT (Hannover) and AQUTE (Bonn). A.~Alberti acknowledges support by the A.~v.~Humboldt Foundation.

\ifwithsupp\papsection A{Molecules in higher dimension}
Generally speaking, a point perturbation has less and less effect in higher dimension. One therefore expects the formation of molecules to become more difficult in higher dimension. This intuition is basically correct, but since  points on a lattice have finite thickness, there is no critical dimension above which molecule formation disappears.

From the analysis given in the paper a sufficient condition is clear: Assume that, for some given total momentum $p=p_0$, the walk in the difference coordinate has a {\it band gap}, i.e., there is a $z$ on the unit circle so that $(W(p_0,k)-z)$ is invertible for all $k$. Choosing $\Ccoll=(\idty-R(z)^{-1})$ (see \eqref{RC}) we then get a walk with an eigenvalue of maximal degeneracy precisely at $z$. By perturbation theory this will typically split into a collection of bound state in a neighborhood of $p_0$. This gives a (possibly small) range of $p$ and suitable initial preparations, which will show the molecule behavior.

The band gap condition can be met by constructing the free walk from a product (not tensor product) of ``slow'' walks in the coordinate directions. Here a slow walk is defined as one with nearly constant eigenvalues $\omega(p)$. For example, consider a walk of the form
\def\twomat#1{\left(\begin{array}{cc}#1\end{array}\right)}
\begin{eqnarray}\label{flatwalk}
    W(p_1,\ldots,p_s)&=&C_0\cdot\twomat{0&e^{ip_1}\\e^{-ip_1}&0}C_1\cdots\\
    &&{}\quad \cdots C_{s-1}\twomat{0&e^{ip_s}\\e^{-ip_s}&0}\cdot C_s.\nonumber
\end{eqnarray}
By choosing appropriate unitary matrices $C_k$, a large variety of $s$-dimensional walks can be implemented. However, when we take all $C_k=\idty$, we get a rather trivial walk which only has steps along the main diagonal and, for all $p$, only the eigenvalues $\pm1$. Then if we take all $C_k$ close to the identity, we get a non-trivial walk in $s$ dimensions, whose eigenvalues are $\omega_+(p_1,\ldots,p_s)\approx 0$ and $\omega_-(p_1,\ldots,p_s)\approx\pi$. The same is true for the non-interacting pair walk $W\otimes W$. Clearly, this now satisfies the band gap condition. Intuitively this construction slows down the group velocity to such an extent that the perturbation at a single point can take effect.

It should be noted, however, that the band gap condition is by no means necessary. For example, if we always take $T$ time steps together, we are looking at the walk $W^T$, whose spectrum, on the one hand, will still show molecule states if $W$ does. On the other hand, any bit of continuous spectrum will be spread out by a factor $T$, and eventually wraps around the circle, maybe with multiplicity $>1$. The molecule states will thus be embedded in the continuum.

In the continuous-time version of the problem \cite{Zoller} the band gap condition is always satisfied , since the lattice kinetic energy is bounded, and a sufficiently strong point interaction can always create a bound state. Hence, as already remarked in \cite{Zoller}, suitably constructed point interactions lead to molecule formation even in high dimension.

\papsection B{Interaction in Quantum Cellular automata}
Unlike the continuous time case, where one can just add Hamiltonian interaction terms, the interaction unitaries in the discrete time case have to be multiplied in some order. This leads to ambiguities for triple collisions. The way to treat these is to consider the many-particle system as a ``lattice gas'', a special kind of quantum cellular automaton(QCA) \cite{Schumacher}. This is analogous to a lattice spin system with the spin up/down states interpreted as occupied/unoccupied. The special feature of a lattice gas (as opposed to a general QCA) is that there is a notion of particle number for each cell, and that the local dynamical rules preserve particle number. The characteristic change of perspective is to focus not so much on the different particles, but on what can happen in a single cell.

One standard transition from single particle walk to a non-interacting lattice gas is Bose/Fermi second quantization. Since for every basis state $\ket{x,\alpha}$ of the one-particle space we have a number operator $n_{x,\alpha}$, we can define the number of particles in the cell $x$ as $\sum_\alpha n_{x,\alpha}$. In the Fermi case this is clearly bounded by $d$, but is unbounded in the Bose case. This is not quite consistent with the idea of finite volume atoms in a potential well, nor with the usual assumption for QCAs that there should be only finitely many states per cell. Indeed the QCA's cell finiteness can be seen as a hard core repulsion condition, so a QCA tends to impose some interaction. There is a procedure for systematically introducing an interaction for a walk which is given in shift\&coin factorization \cite{Holger}. We illustrate it for the factorization \eqref{hadamard} of the Hadamard walk.

The QCA system consists of a double chain of qubits, which we may call the right moving chain and the left moving chain. This describes the action of the shift operation: In the shift step the entire right (resp.\ left) moving chain is shifted on step to the right (resp.\ left). The coin operation acts at each site. We denote the basis states for all qubits by $\ket0$ for empty and $\ket1$ for occupied, so at each site we have the basis vectors $\ket{\alpha\beta}=\ket\alpha_{\rm left}\otimes\ket\beta_{\rm right}$. In order to guarantee particle number conservation, the coin has to act in block diagonal form on the number subspaces. The empty space $\ket{00}$ will just be left invariant. There is no loss of generality in this choice, because we are free to choose a global phase. On the basis vectors $\ket{01}$ and $\ket{10}$ of the one particle space we just use the coin operation of the free walk. Then it is easy to see that the subspace of total particle number $1$ is invariant, and isomorphic to the free one-particle walk. The only free parameter in this construction is the phase $\gamma$ for the two-particle state $\ket{11}$. Again, it is easy to check that on the $2$-particle subspace we exactly get the singlet collision case discussed in the paper. However, the construction also extends to arbitrarily many particles.

This construction can be extended to other walks with an explicit decomposition into shift and coin operations (allowing also for many factors). Since such decompositions are not unique there are some trade-offs possible\cite{Holger}: For example, one can use a decomposition in which only one if the $d$ states is ever shifted. The multiple chain picture then only has a qubit chain for transport, and a stationary chain, with an empty state and, for particle number $1$, the remaining $(d-1)$ states of the one-particle walk. This leads to a QCA with $2d$ states per cell. However, the interaction neighborhood size of this QCA may be much larger than that of the one-particle walk. In contrast, taking a moving chain for every one of the $d$ states suggests a picture with $2^d$ states per cell, and in this case (at least in 1D) the neighborhood can be arranged to be the same as for the walk. Further options are introduced by allowing more than single occupation of the one-particle coin states, as in the free Bose walk.

Which of these many possibilities is realized is, of course, a question that has to be answered by analyzing the physics of the given system. The point to be made here is only that the language of QCAs makes clear what physical information is required to specify an ``interaction'' completely.

\papsection C{A Unitarity Lemma}
In this appendix we prove the claim that the operator $(\idty-R(z)^{-1})$ appearing in \eqref{RC} is unitary. We will do this by proving the following, more general claim:

Consider a unitary operator $W$ on a Hilbert space $\HH$ and a point $z$  on the unit circle ($z\in\Cx$, $\abs z=1$) not in the spectrum of $W$, which just means that the inverse $(W-z\idty)^{-1}$ exists and is a bounded operator. Now let $N$ be an arbitrary projection and set $R(z)=N(W-z\idty)^{-1}WN$, considered as an operator on $N\HH$. We claim that this operator is invertible and that $U=\idty-R(z)^{-1}$ is unitary.

The crucial step for the proof is to show the following relation
\begin{equation}\label{r+rstar}
    R(z)+R(z)^\dagger=\idty.
\end{equation}
We can show this for the special case $N=\idty$, and then multiply the above equation with $N$ from both sides. Since $W$ is unitary it has a spectral decomposition $W=\int\!\Omega\ E_W(d\Omega)$, where $E_W$ is the spectral measure on the unit circle. We can then express the operator \eqref{r+rstar} directly in the functional calculus (we use $\dagger$ for hermitian adjoint and an overbar for complex conjugate):
\begin{eqnarray}
    R(z)+R(z)^\dagger
       &=& \int\left(\frac\Omega{\Omega-z}+ \frac{\overline\Omega}{\overline\Omega-z}\right)E_W(d\Omega)
            \nonumber\\
       &=&  \int\frac{\Omega({\overline\Omega}-z)+ {\overline\Omega}(\Omega-z)}
                    {\abs{\Omega-z}^2}\,E_W(d\Omega)
            \nonumber\\
       &=&  \int\frac{2 -2 \Re({\overline\Omega}z)}
                    {2 -2 \Re({\overline\Omega}z)}\,E_W(d\Omega)
             \nonumber\\
       &=& \idty\nonumber
\end{eqnarray}
This determines the hermitian part of $R(z)$, and we can write $R(z)=\frac12(\idty+iK)$ for some hermitian operator $K$ on $N\HH$.

For the next we use the functional calculus of $K=\int k\,E_K(dk)$, with the spectral measure of $K$, which is now supported on the real axis. We note that due to the projection $N$, which need not commute with $W$, this measure cannot be easily obtained from the spectral measure $E_W$. Then
\begin{eqnarray*}
    U &=& \idty- R(z)^{-1}
           \noindent\\
      &=&\int\left(1-\frac1{\frac12(1+ik)}\right)\,E_K(dk)
           \noindent\\
      &=&\int\frac{-1+ik}{1+ik}\,E_K(dk).
\end{eqnarray*}
Now, since $\abs{-1+ik}=\abs{1+ik}$ the integrand has absolute value $1$, and so the integral represents a unitary operator.

\papsection D{Overlap with bound states}
The time evolution of a two-particle quantum state under the considered quantum walk $W$ is essentially determined from its overlap with the bound states $\Psi(p)$ in Fourier space. This is also important for experimental implementations since it describes how efficient some preparation, e.g., ``both particles in singlet state at the origin'', is for generating molecules. In other words, we want to estimate the splitting in Fig.~\ref{fig:positions} between the total probability near the diagonal and the rest, i.e., the probability for ballistically moving away from the diagonal. Since the total  momentum $p
$ is a conserved quantity we omit the $p$-dependence of $W$ in the following discussion.

First, let us calculate the normalization of the eigenstate $\Psi$ of the quantum walk corresponding to eigenvalue $z$ as determined from \eqref{resolve}. Abbreviating $\psi_\Gamma = (\idty-\Ccoll)\psi$ we get
\begin{align}\label{es_norm}
	\norm{\Psi}^2 &= \int \frac{d^s k}{2 \pi} \norm{(W(k)-  z)^{-1}\psi_\Gamma}^2\nonumber\\ & = \braket{\psi_\Gamma}{\left(-z \int \frac{d^s k}{(2 \pi)^s}\frac{ W(k)}{(W(k)-z)^2}\right)\psi_\Gamma}\nonumber\\ &= \braket{\psi_\Gamma}{-z \frac{d}{dz} R(z) \psi_\Gamma}
\end{align}
where we used the definition of $R(z)$ from equation \eqref{rr} and the unitarity of $W(k)$. This yields now the probability to capture the walker initially prepared in an internal state $\Phi$ located at the origin in the bound state
\begin{align}\label{cap_prob}
  P_{cap}= \frac{\abs{\braket{\Phi}{R(z)\psi_\Gamma}}^2}{\norm{\Psi}^2} =  \frac{\abs{\braket{\Phi}{\psi}}^2}{\norm{\Psi}^2}\;,
\end{align}
since $\psi$ respects \eqref{RC}. We want to evaluate \eqref{cap_prob} for our standard example, the Hadamard Walk with singlet collisions and the singlet state $\psi_{-}$ as the initial state, where $\psi_\Gamma$ is given by $(1-\gamma)\psi_{-}$. By \eqref{es_norm}  we have to calculate the reciprocal of
\begin{align*}
 \norm{\Psi}^2&=  -\abs{1-\gamma}^2\, z \frac{d}{dz}\braket{\psi_{-}}{R(z) \psi_{-}}  =-i \gamma^{-1} \frac{d\gamma}{d\omega}
\end{align*}
where we used \eqref{RC} and the convention $z=e^{i\omega}$ in the last step. Defining $\eta$ via $e^{i\eta}=-\frac{z}{\gamma}$ one can deduce from \eqref{omegap} the relation
\begin{align}\label{eq:eta}
  \cos \eta  = \cos p - 2\cos \omega\;.
\end{align}
Hence, we get for the capture probability $P_{cap}$ in the bound state corresponding to the eigenvalue $z$ the expression
\begin{align}\label{eq:Pcap}
  P_{cap} = \left(1+2 \,\frac{\sin\omega}{\sin\eta}\right)^{-1} \,.
\end{align}
Note that this is a function of the total momentum $p$ since $\eta$ as well as $\omega$ depend on it. Analyzing this dependency, we see that as $\omega$ approaches a band edge, $\sin\eta$ tends to zero and therefore such $p$ values are suppressed as Fourier components of the bound state. That behavior can also be seen in Fig. \ref{fig:bound}.

\begin{figure}[ht]
  \includegraphics[width=8.2cm]{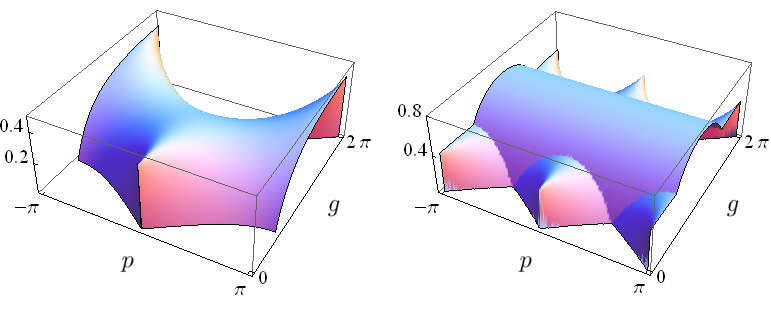}
  \caption{The figure shows the capture probability $P_{cap}$ for the $\omega_+$ branch (left panel) and the sum of both branches (right panel) as a function of the total momentum $p$ and of the interaction phase $\gamma=e^{ig}$. }
  \label{fig:bound}
\end{figure}

For fixed interaction phase $\gamma=e^{ig}$ the ratio between the occurrence of molecules and unbound two particle states in an experiment is determined from the integral of $P_{cap}$ with respect to the total momentum $p$. As can be seen from Fig. \ref{fig:darthvader} there is a fairly high probability to observe molecules with a maximum of $2/3$ at $g=\pi$.

\begin{figure}[ht]
  \includegraphics[width=4.8cm]{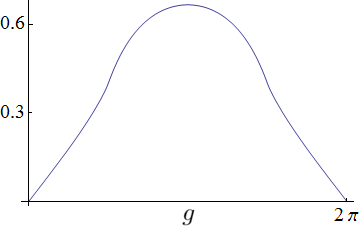}
  \caption{The integrated capture probability $P_{cap}$ is plotted over the interaction phase $\gamma=e^{ig}$. It reaches its maximal value of $2/3$ at $g=\pi$.}
  \label{fig:darthvader}
\end{figure}

Expression \eqref{eq:Pcap} also allows us to study the long time behavior of the molecule's position distribution scaled by $1/t$, where $t$ is the discrete time parameter. As outlined in corollary 7 in \cite{timerandom}, this asymptotic distribution can be evaluated with knowledge of the group velocity \eqref{velo} and the capture probability \eqref {eq:Pcap}. The result is shown in Fig. \ref{fig:AsyDist}.

\begin{figure}[ht]
  \includegraphics[width=8.2cm]{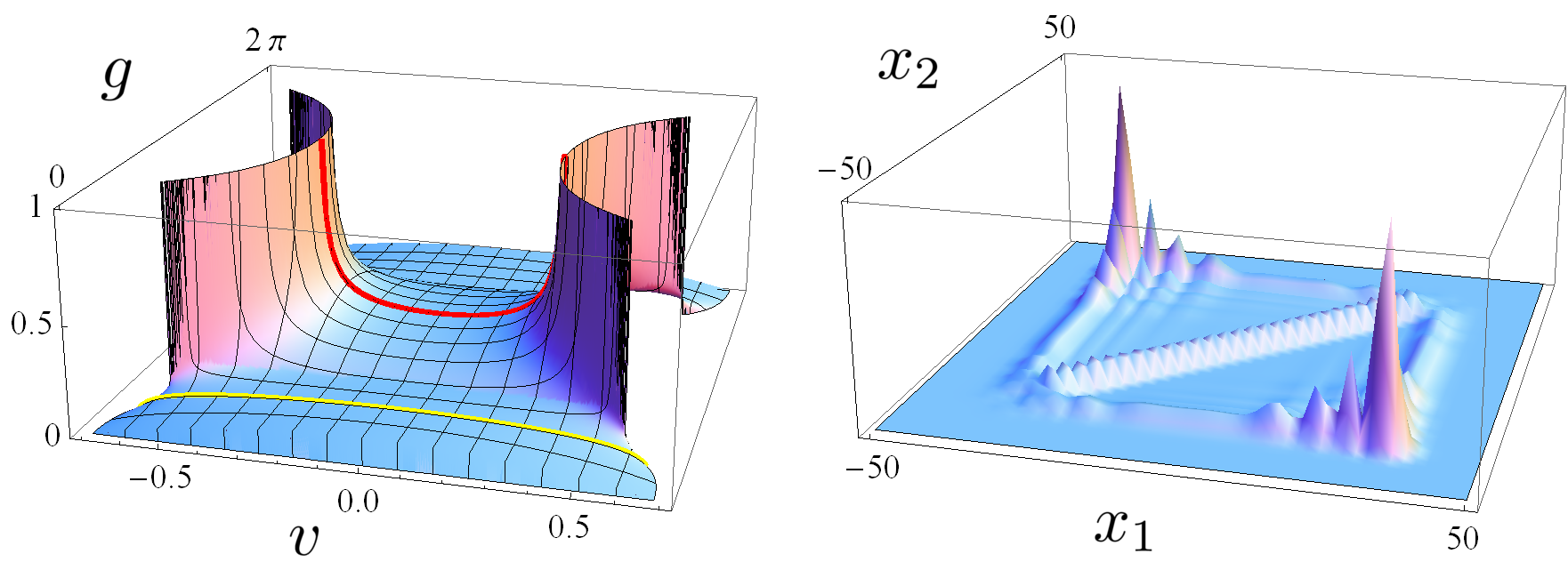}
  \caption{The left panel shows the molecule's asymptotic position distribution depending on the interaction phase $\gamma=e^{ig}$. The yellow line corresponds to an interaction phase of $g=0.8$, the red line to an interaction phase of $g=\pi$. The flat distribution profile for $g=0.8$ matches the flat position distribution of the bound state shown in the right panel for $t=50$ time steps of the interacting quantum walk. The two peaked caustic behavior in the asymptotic position distribution for $g=\pi$ can be compared to the corresponding panel in Fig. \ref{fig:positions}.}
  \label{fig:AsyDist}
\end{figure}

\papsection E{Analytic form of bound states}
The goal of this section is to derive an explicit formula for the bound states of a quantum walk with a point defect at the origin. We then apply this procedure to our standard example, the Hadamard walk with singlet collisions. This allows us to derive an explicit form of the $p$ dependent transformation needed to construct the virtual walk \eqref{mwalkcoin} of the molecule as explained in the main part of the paper.

The components $\Psi_x$ of the unnormalized eigenvector $\Psi$ corresponding to lattice site $x$ can be determined from \eqref{resolve}, with the substitution $v=e^{ik}$ the formula reads
\begin{eqnarray}
\label{PsiEq}
\Psi_x &=&\frac1{2\pi i}\int\!\!\frac{dv}{v^{x+1}}\ (W(v)-z)^{-1}W(v)\psi_\Gamma \nonumber\\
&=&\frac1{2\pi i}\int\!\!\frac{dv}{v^{x+1}}\left(\idty +z(W(v)-z)^{-1}\right) \psi_\Gamma \, .
\end{eqnarray}
Here, we have to substitute the correct form \eqref{omegap} of the eigenvalue $z=e^{i\omega}$. The initial state in our example is the singlet state $\psi_-$ which effectively reduces the problem to the Fermi sector. Of course, the Fermi symmetry is reflected in the eigenstate $\Psi$. If we denote the operator which exchanges tensor factors by $\FF$, that is, $\FF \eta \otimes \chi =\chi\otimes \eta$ for all vectors $\eta ,\chi \in\Cx^2$, then $\FF W(v) \FF=W(v^{-1})$ and hence $\Psi_x=-\FF\Psi_{-x}$ for all $x\in\Ir$.

In order to determine $\Psi_x$ for negative $x$ we use residual calculus which requires knowledge of the poles of the integral kernel in \eqref{PsiEq}. The operator $(W(v)-z)^{-1}$ may have singularities at the zeros of the quadratic polynomial $p(v)=v\det (W(v)-z)$ and possibly at $v=0$. Conjugation of $W(v)-z$ by $\FF$ yields the equality $\det (W(v)-z)=\det (W(v^{-1})-z)$, which implies that the singularities of $(W-z)^{-1}$ are either of first order and inverses with respect to each other or $p(v)$ is constant, in which case the operator $(W-z)^{-1}$ has a pole of first order at $v=0$. Analyzing the case where $p(v)$ is not constant we get the following expression for the singularity with modulus smaller than one, denoted by $v_1$,
\begin{eqnarray}
    v_1&=&-\cos \eta -\cot \frac{g}{2} \sin \eta
                   \nonumber\\
               && \mskip-30mu\mbox{ provided}\ \sin(\eta+g)\cdot\sin(\eta)<0\,,
\end{eqnarray}
with $\eta$ as in \eqref{eq:eta}. Clearly, $\Psi_0=\psi_-$ and if $x<0$ only $z(W(v)-z)^{-1}$ contributes to the integral \eqref{PsiEq}. Since the integral kernel is analytic at $v=0$ it remains to calculate the component wise residual operator
\[
R_{v_1}=\mathop{{\mathrm{Res}}}\limits_{v \rightarrow v_1}\left( v^{-1}z(W(v)-z)^{-1}\right)
\]
at $v_1$. The coefficients of the normalized eigenstate $\tilde\Psi$ in terms of this operator read
\[
\tilde\Psi_x =P_{cap}\cdot\left\{
\begin{array}{lll}
(1-\gamma )v_1^{|x|}R_{v_1}\psi_- &,& x<0\\
\psi_- &,& x=0\\
-(1-\gamma )v_1^{|x|}\FF R_{v_1}\psi_- &,& x>0
\end{array}
\right.\,.
\]
Note that since the modulus of $v_1$ is smaller than one the state $\tilde\Psi$ decays exponentially in $\vert x\vert$. This implies that the distance of the two particles in the molecule state is exponentially concentrated around zero.

In the case that the singularities of $(W(v)-z)^{-1}$ are at $v_1=0$ the eigenstate $\Psi$ is strictly localized on a finite set of lattice sites. This is because for $v_1=0$ the integral kernel in \eqref{PsiEq} is analytic if $\vert x\vert>1$, hence $\Psi_x=0$ for these cases. For our example this happens exactly at $p=\pm \omega(p)$. This property leads to an interesting feature of the considered quantum walk: By engineering the initial state of the quantum walk on few lattice sites it is possible to generate states which are close to a momentum eigenstate at points $p=\pm\omega(p)$. Here, the spread of the initial state in the relative coordinate is bounded by the spread of the eigenstate $\tilde\Psi_x$ and in the center of mass direction by the desired accuracy of the momentum preparation.

With the help of the eigenstates $\tilde\Psi_\pm$ corresponding to the two branches $\omega_\pm$ we can construct the one dimensional quantum walk mimicking the time evolution of the molecule explicitly. This is done by identifying the states $\tilde\Psi_\pm$ with the $p$ dependent eigenvectors of some one dimensional quantum walk with the same dispersion relation $\omega_\pm$.

\papsection F{Fast Molecules}
So far only the two particle Hadamard Walk with singlet collisions was considered as explicit exemplification of the ideas of this paper. From Fig. \ref{fig:positions} and Fig. \ref{fig:AsyDist} one could think that the bound states are slower or at most as fast as the free walkers.

In this section we will give two numerical examples to show that this is not true in general and that it is possible to design interactions that allow for molecules spreading faster then the free walk. Since in the case of singlet collisions the only freedom lies in the selection of the interaction phase $\gamma$ we turn to the Bose case where we can choose an arbitrary three by three unitary matrix in symmetric subspace.

A trivial possibility to generate such fast molecules is choose an interaction matrix that counteracts the coin operation. In that case the walk operator for the two particle Hadamard walk is given by
\begin{align*}
   W_\Ccoll&=(W_1\otimes W_1)\bigl( (\idty-N)+ H\otimes H N\bigr) \\
  & =(S\otimes S)(H\otimes H (\idty-N) + \idty N)\; ,
\end{align*}
which acts on the walkers only with the shift at the collision point. If we now prepare the particles at the same lattice side for example in the state $\ket{\hspace{-3pt}\uparrow\uparrow}$ they will be just shifted to the left with maximal velocity.

\begin{figure}[ht]
  \includegraphics[width=8.2cm]{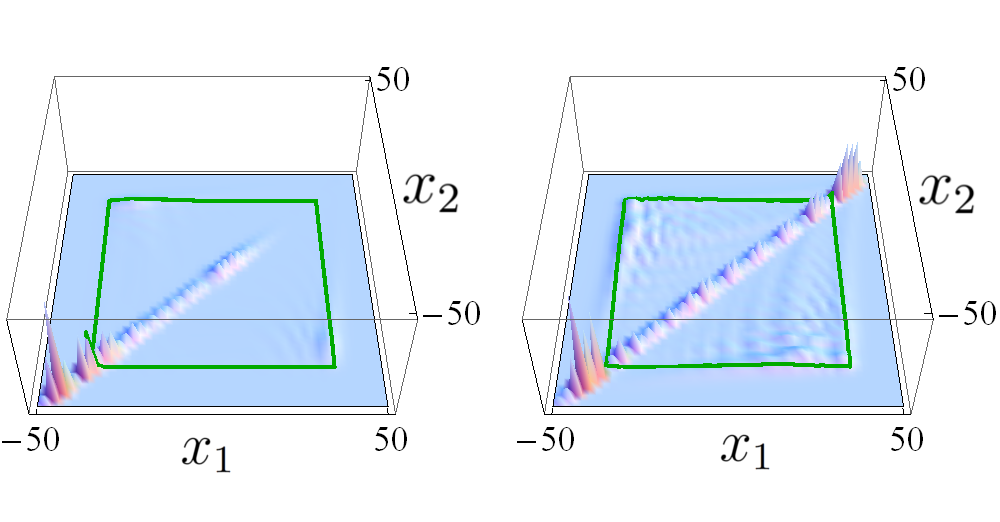}
  \caption{The two panels show two particle Hadamard walks with different bosonic interactions. In both cases the molecule state extends beyond the green square indicating the maximal velocity of the free Hadamard walk.}
  \label{fig:fastMol}
\end{figure}

Two nontrivial examples of fast molecules are given in Fig. \ref{fig:fastMol}, were we have choosen two  unitary interactions for which the bound state in the symmetric subspace spreads faster than the individual particles of the free quantum walk. Moreover it is even possible to design the interaction in such a way that the resulting molecule travels nearly at the maximal possible velocity as in the trivial case.
\fi

\bibliography{moleculit}

\end{document}